\documentclass[aps,prd,preprint,superscriptaddress,tightenlines,nofootinbib,showpacs]{revtex4}

\usepackage{graphicx}
\usepackage{dcolumn}
\usepackage{bm}

\begin{document}

\preprint{CLNS 06/1976}       
\preprint{CLEO 06-18}         
\def\etaP{\eta^{\prime}}

\title{$\chi_{c0}$ and $\chi_{c2}$ Decays into 
       $\eta\eta$, $\eta\etaP$, and $\eta^{\prime}\eta^{\prime}$
Final States}

\author{G.~S.~Adams}
\author{M.~Anderson}
\author{J.~P.~Cummings}
\author{I.~Danko}
\author{J.~Napolitano}
\affiliation{Rensselaer Polytechnic Institute, Troy, New York 12180}
\author{Q.~He}
\author{J.~Insler}
\author{H.~Muramatsu}
\author{C.~S.~Park}
\author{E.~H.~Thorndike}
\author{F.~Yang}
\affiliation{University of Rochester, Rochester, New York 14627}
% ...
\collaboration{CLEO Collaboration} 
\noaffiliation

\noaffiliation

\date{\today}

\begin{abstract}
Using a sample of $3 \times 10^6$ $\psi(2S)$ decays collected
by the CLEO~III and CLEO--c detector configurations, we present
results of a study of $\chi_{c0}$ and $\chi_{c2}$ decays
into $\eta\eta$, $\eta\etaP$, and $\etaP\etaP$ final states. 
We find $B(\chi_{c0}\to\eta\eta)= (0.31\pm0.05\pm0.04\pm0.02)$\%, 
$B(\chi_{c0}\to\eta\etaP) 
<0.05$\% at the 90\% confidence level, and
$B(\chi_{c0}\to\etaP\etaP) = (0.17\pm0.04\pm0.02\pm0.01)$\%. 
We also present upper limits
for the decays of $\chi_{c2}$ into these final states. 
These results give information
on the decay mechanism of $\chi_c$ states
into pseudoscalars.

\end{abstract}

\pacs{13.25.Gv, 14.40.Gx}
\maketitle


In the standard quark model, the $\chi_{cJ}$ ($J=0,1,2$), 
mesons are $c\bar{c}$ states
in an $L=1$ configuration.
As they cannot be produced directly in $e^+e^-$ collisions, they are 
less well studied than the $\psi$ states. However, 
$\psi(2S)\to\gamma\chi_{cJ}$
decays yield many $\chi_{cJ}$ mesons, and $e^+e^-\to\psi(2S)$ is a very clean  
environment for $\chi_{cJ}$ investigation. 
In this paper, we concentrate on the 
two-body decays of the $\chi_{c0}$ and $\chi_{c2}$ into $\eta\eta$, 
$\eta\etaP$,
and $\etaP\etaP$ final states\footnote{We do not consider $\chi_{c1}$ decays into these
final states, as they are forbidden by simple spin-parity conservation.}.
Knowledge from these decay rates leads to information on the quark and gluon
nature of both the $\chi_c$ parents and their pseudo-scalar daughters, and a 
greater understanding of the decay mechanisms of the $\chi_c$ mesons \cite{ZHOU}. 

The data presented here were taken by the CLEO detector operating at the Cornell 
Electron Storage Ring with a center of mass energy corresponding to the $\psi(2S)$
mass of 3.686 $\rm{GeV/c^2}$. The data were taken with two different detector 
configurations.
An integrated luminosity of 2.74 ${\rm pb^{-1}}$ was collected
using the CLEO III detector configuration \cite{CLEOIII}, 
and 2.89 ${\rm pb^{-1}}$ using the CLEO-c detector configuration \cite{CLEO-c}.
The total number of $\psi(2S)$ events is calculated as $3.08\times 10^6$,
determined according to the
method described in \cite{PSI2S}.
The vital detector component for this analysis, the CsI crystal calorimeter \cite{CLEOII}, is
common to the two configurations and has an energy resolution of
2.2\% at 1 GeV, and 4.0\% at 100 MeV.  

We detect our $\eta$ candidates in the decay modes 
$\gamma\gamma$, $\pi^+\pi^-\pi^0$,
and $\pi^+\pi^-\gamma$, and our $\etaP$ 
candidates in the modes 
$\gamma\pi^+\pi^-$ and $\eta\pi^+\pi^-$, where
the $\eta$ is reconstructed 
using the $\eta$ decay modes previously listed. 
We
look for all combinations of distinct $\eta$ and $\etaP$ candidates,
with the exception of the 
case where both $\eta$ candidates in the event
decayed to $\gamma\gamma$. This last case was excluded because 
its detection depends on the operation of an all-neutral trigger 
whose efficiency varied over the running period because of hardware changes.
All our final states thus 
contain at least two charged particles, 
and the trigger
efficiency for these events is close to 100\% after all other requirements.

We define photon candidates as clusters in the CsI having a
shower shape
consistent
with being due to a photon. 
We form $\eta\to\gamma\gamma$ candidates,
from pairs of photon candidates, using the event
vertex found from the charged tracks 
as the position of origin of the photons. 
We kinematically constrain the two-photon combination
to the known $\eta$ mass, and those with
a $\chi^2 < 10$ (for one degree of freedom) for this fit are
retained 
for further analysis. 
The same procedure is used for $\pi^0$ candidates. We then 
proceed to make $\eta\to\pi^+\pi^-\pi^0$ candidates by constraining the 
$\pi^+\pi^-\pi^0$ combinations to the nominal
$\eta$ mass and again requiring a $\chi^2$ of less than
10 for one degree of freedom. 
Similarly $\etaP$ candidates are built, either from the $\eta\pi^+\pi^-$
combinations, or from combining $\pi^+\pi^-$ and $\gamma$ candidates 
(with $E_{\gamma}> $50 MeV), mass constraining
the resultant 4-momentum to the $\etaP$ mass and 
requiring them to have a $\chi^2 < 10$ for the one degree of freedom
of this constraint. 
The mass resolution of the reconstructed $\eta\to\gamma\gamma$ combinations
is around 12 ${\rm MeV/c^2}$ before the mass constraint, 
and the analogous numbers for
$\eta \to \pi^+ \pi^- \pi^0$, $\etaP \to \gamma \pi^+\pi^-$, 
and $\etaP \to \eta\pi^+\pi^-$ are 3 ${\rm MeV/c^2}$, 7 ${\rm MeV/c^2}$, 
and 2.5 ${\rm MeV/c^2}$ respectively.

If we have two distinct $\eta^{(\prime)}$ candidates, 
we combine them into a 
$\chi_{c}$ candidate. At this stage of the analysis, the signal to noise
ratio is poor, and the invariant 
mass resolution of the $\chi_c$ is around 15 ${\rm MeV/c^2}$.
We then search for any unused photon in the event, 
and add that to the $\chi_c$ candidate to make
a $\psi(2S)$ candidate. This $\psi(2S)$ is then kinematically 
constrained to the 4-momentum of the beam, 
the energy of which is taken as the known
$\psi(2S)$ mass, and the momentum is non-zero only by a tiny amount due to the
finite crossing angle ($\approx 2$ mrad per beam) in CESR. 
To make our final selection, 
we require the $\psi(2S)$ candidate to have a  
$\chi^2$ of less than 25 for the 4 degrees of freedom for this fit; 
this cut rejects many background combinations. 
This kinematic fit greatly improves the mass resolution of the 
$\chi_c$ candidate.

To study the efficiency and resolutions, 
we generated Monte Carlo samples
for each $\chi_c$ state into each final state, 
using a GEANT-based detector simulation\cite{GEANT}.
The simulated events have a distribution of
$(1+\lambda \cos^2 \theta)$, where $\theta$ is the 
radiated photon angle relative to the
positron beam direction, and $\lambda=$ 1 for the 
$\chi_{c0}$ and $\lambda=$ 1/13 for the $\chi_{c2}$, 
in accordance with expectations for an E1 transition.
The mass resolution and efficiencies are shown in Table~I. 
The resolutions
varied a little with the 
detector configuration and decay chain, but are approximated
by single Gaussian functions. The efficiencies 
shown include all the relevant 
branching fractions of the $\eta$ and $\etaP$. 
\begin{table}[htb]
\caption{Efficiencies (in \%) and resolutions (in ${\rm MeV/c^2}$) obtained
from analysis of Monte Carlo generated events.}

\begin{tabular}{|l|c|c|c|c|c|c|c|}
\hline
Mode  & \multicolumn{2}{c|}{ $\chi_{c0}$ } & \multicolumn{2}{c|}{ $\chi_{c2}$} \\
\hline
      & Efficiency & Resolution & Efficiency & Resolution \\
&           \%   & ${\rm MeV/c^2}$ & \% & ${\rm MeV/c^2}$ \\
 \hline
$\eta \eta$       & 5.46 & 7.6 & 5.60 & 6.3   \\
$\eta \etaP$      & 5.47 & 7.0 & 5.47 & 5.8   \\
$\etaP \etaP$     & 4.69 & 5.4 & 4.67 & 4.9   \\
\hline
\end{tabular}
\end{table}
                                                                                    
The final mass plots are shown in Fig.\ 1. Clear peaks are found for the 
decays $\chi_{c0}\to\eta\eta$ and $\chi_{c0}\to\etaP\etaP$, but the other 
four decays under consideration show no significant signals.
These plots are each fit with two signal shapes comprising Breit-Wigner
functions convolved
with Gaussian resolutions, together with a flat background term. 
The masses and widths
of the Breit-Wigner functions 
were fixed at their standard values \cite{PDG}, and 
the widths of the Gaussian resolution functions 
fixed at the values shown in 
Table~I. We find signals of $47.8\pm7.7$ events for
$\chi_{c0}\to\eta\eta$,
and $22.7\pm5.3$ events for $\chi_{c0}\to\etaP\etaP$. 
To find the limits on the number of events for each of the 
other decay modes,
we plot the probability function for a series of different signal 
yields, and place the limit so that 90\% of the physically allowed 
region is below its value.
The signal yields and limits are summarized in Table~II.
Our detected signal events in $\chi_{c0}$ decays have a 
$\theta$ angle distribution consistent with the value of $\lambda=1$
as defined above. They also make a uniform band of events in the 
$M^2(\gamma\eta_1)$ versus $M^2(\gamma\eta_2)$ Dalitz plot, as expected
for the decay of a spin-0 particle.

To convert the yields to branching fractions, we divide by the 
number of $\psi(2S)$ events in the data sample multiplied by
the detector efficiency
and by the 
branching fractions for $\psi(2S)$ into $\chi_{cJ}$, for which 
we use the CLEO measurements of $B(\psi(2S)\to\gamma\chi_{c0})=
9.22\pm0.11\pm0.46$\%, and  $B(\psi(2S)\to\gamma\chi_{c2})=
9.33\pm0.14\pm0.61$\% \cite{PSI2S}.

The systematic uncertainties in the branching 
fractions are summarized in Table~III. 
The systematic uncertainties
due to fitting the peaks were evaluated by floating,
in turn, the mass, 
instrinsic width, and
Gaussian resolution width. The maximum change in yield for our
highest-statistics signal 
($\chi_{c0}\to\eta\eta$) from this process gives
us our measure of this systematic uncertainty, 5\%.
The requirement on the $\chi^2$ of the
constraint to the beam 4-momentum has been checked by changing 
the cut and noting
the change in the yield. 
This has been done for the modes under investigation here,
and also in other modes such as $\eta\pi^+\pi^-$ that have higher statistics. 
Based on
this study we place a systematic uncertainty of 3.5\% on the efficiency 
of this 
requirement. The uncertainties due to track reconstruction are small
and well understood. The biggest systematic uncertainty is that due 
to the photon reconstruction, 
which is set at 2\% per photon.
This was derived from a series of studies of photon reconstruction in well 
understood decays such as $\psi(2S)\to J/\psi\pi^0\pi^0$. 
For the two signals $\chi_{c0}\to\eta\eta$ and $\chi_{c2}\to\eta\etaP$ 
we add the systematic uncertainties in quadrature, except for the 
one due to the $\psi(2S)\to\gamma\chi_c$ branching fractions which we keep
separate, when calculating the final branching
fractions. 
For evaluating the limits in the cases where there is no
significant signal, we take the probability density function, and 
divide 
each entry by the efficiency smeared by the total systematic uncertainty, and 
find the branching fraction that includes 90\% of the total area.
A rough check on our ability to find the efficiency of the entire decay
chain is possible by measuring the number of events consistent with the
decay chain $\psi(2S)\to\eta J/\psi, J\psi\to\gamma\etaP$. This has 
the same final state particles as $\psi(2S)\to\gamma\chi_c,\chi_c\to\eta\etaP$.
Using the Review of Particle Properties\cite{PDG} 
values of the branching fractions, we expect to see
$19.2\pm 2.4$ events in this mode, and we observe $20.8\pm 4.7$, which is 
consistent with this expectation.

\begin{table}[htb]
\caption{Signal yields and branching fraction results  for each decay mode.
The uncertainties are statistical, systematic due to this measurement, and systematic
due to the $\psi(2S)\to\chi_{cJ}$, respectively. The limits on the branching fractions
include all systematic uncertainties.}

\begin{tabular}{|l|c|c|c|c|c|c|c|}
\hline
Mode  & \multicolumn{2}{c|}{ $\chi_{c0}$ } & \multicolumn{2}{c|}{ $\chi_{c2}$} \\
\hline
      & Yield & B.F. (\%) & Yield & B.F. (\%) \\
\hline
$\eta \eta$       & $47.8\pm7.7$ & $0.31\pm0.05\pm0.04\pm0.02$ & 
                      $<7.2$ & $<0.047$   \\
$\eta \etaP$      & $<7.5$ & $<0.050$  & $<3.4$ & $<0.023$   \\
$\etaP \etaP$     & $22.7 \pm 5.3$ & $0.17\pm0.04\pm0.02\pm0.01$ & $<4.0 $ & $<0.031$   \\
\hline
\end{tabular}
\end{table}

\begin{table}[htb]
\caption{Systematic uncertainties expressed in percent. }

\begin{tabular}{|l|c|c|c|c|c|c|c|c|}
\hline
Mode & $N_{\psi(2S)}$ & Trig. & Fit & Tracks & $\gamma$ & MC stats & Cuts & Total  \\
\hline
$\eta\eta$                     & 3 & 1 & 5 & 1.4 & 10 & 3 & 3.5 & 12  \\
$\eta\eta^{\prime}$            & 3 & 1 & 5 & 2.1 & 8 & 3 & 3.5 & 11  \\
$\eta^{\prime}\eta^{\prime}$   & 3 & 1 & 5 & 2.8 & 6 & 3 & 3.5 & 10  \\
\hline
\end{tabular}
\end{table}

Our results can be interpreted in the model of Qiang Zhao \cite{ZHOU}, who predicts
these decay rates as a function of the QCD parameter, $r$, which is the ratio of
doubly- to singly-OZI-suppressed decay diagrams. The measurements are 
consistent with small values of $r$, which indicates that it is the singly
suppressed OZI diagram that dominates these decays. We also present
a limit on the branching fraction for $\chi_{c2}$ decays to $\eta\eta$ that is
tighter than that of BES \cite{BES}, and the first limits 
for $\chi_{c2}$ decays into $\eta\etaP$ and $\etaP\etaP$. The
limit for $\chi_{c2}\to\eta\eta$ is slightly below the theoretically 
predicted branching fraction for $r=0$, but all 
three limits are 
consistent with $r$ being small and negative. More definitive conclusions
require a larger sample of $\chi_{c2}$ decays.

In summary, 
we find a branching fraction for $\chi_{c0}\to\eta\eta$ of $(0.31\pm0.05\pm0.04\pm0.02)$\%. 
This is larger than, but consistent with, two lower-statistics measurements of 
BES \cite{BES} and E-835 \cite{E835}. 
We make the first measurement of $\chi_{c0}\to\etaP\etaP$ of 
$(0.17\pm0.04\pm0.02\pm0.01)$\%. We find no 
signal for $\chi_{c0}\to\eta\etaP$ 
and are able to set an upper 
limit on this branching fraction of $<0.05$\%
at the 90\% confidence level. We also set limits of the branching fraction for
$\chi_{c2}$ decaying to $\eta\eta$, $\eta\etaP$, and $\etaP\etaP$ of
0.047\%, 0.023\% and 0.031\% respectively. 
Our results imply that 
singly OZI-suppressed diagrams dominate in these decays.

\begin{figure}[htb]

\caption{Invariant mass distributions for (a) $\eta\eta$, 
(b) $\eta\etaP$, and (c) $\etaP\etaP$.  
The fits are described in the text.
}
\includegraphics*[width=3.4in]{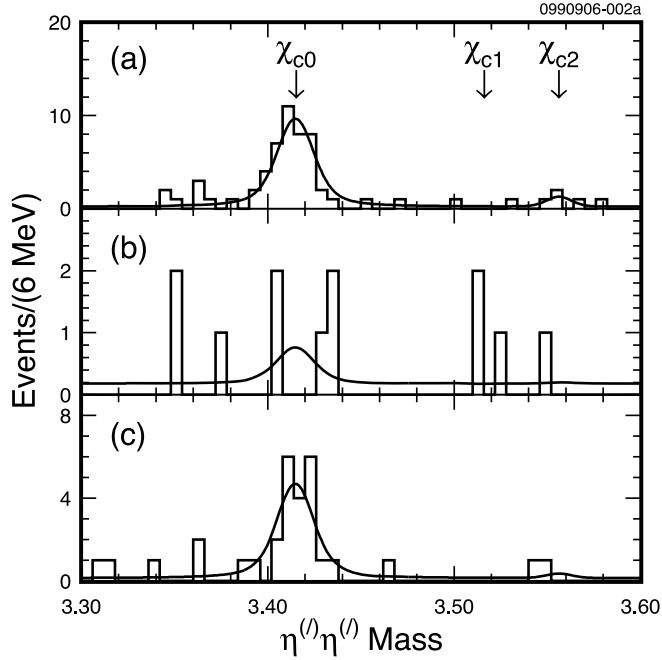}

\end{figure}

We gratefully acknowledge the effort of the CESR staff 
in providing us with excellent luminosity and running conditions. 
D.~Cronin-Hennessy and A.~Ryd thank the A.P.~Sloan Foundation. 
This work was supported by the National Science Foundation,
the U.S. Department of Energy, and 
the Natural Sciences and Engineering Research Council of Canada.

\end{document}